# Progresses and Challenges in Link Prediction

Tao Zhou

CompleX Lab, University of Electronic Science and Technology of China, Chengdu 611731,

People's Republic of China.

## SUMMARY

Link prediction is a paradigmatic problem in network science, which aims at estimating the existence likelihoods of nonobserved links, based on known topology. After a brief introduction of the standard problem and metrics of link prediction, this review will summarize representative progresses about local similarity indices, link predictability, network embedding, matrix completion, ensemble learning and some others, mainly extracted from related publications in the last decade. Finally, this review will outline some long-standing challenges for future studies.

## INTRODUCTION

Network is a natural and powerful tool to characterize a huge number of social, biological and information systems that are consisted of interacting elements, and network science is currently one of the most active interdisciplinary research domains [1][2]. Link prediction is a paradigmatic problem in network science that attempts to uncover missing links or predict future links [3], which has already found many theoretical and practical applications, such as reconstruction of networks [4][5], evaluation of evolving models [6][7], inference of biological interactions [8][9], online recommendation of friends and products [10][11], and so on.

Thanks to a few pioneering works [12][13][14][15], link prediction has been one of the most active research domains in network science. Early contributions were already summarized by a well-known survey article [3], and this review will first define the standard problem and discuss some well-known evaluation metrics, and then introduce most representative achievements in the last decade (mostly published after [3]), including local similarity indices, link predictability, network embedding, matrix completion, ensemble learning and some others. Lastly, this review will show limitations of existing studies as well as open challenges for future studies.

## EVALUATION

Consider a simple network $G(V, E)$, where $V$ and $E$ are sets of nodes and links, the directionalities and weights of links are ignored, and multiple links and self-connections are not allowed. We assume that there are some missing links or future links in the set of nonobserved links $U \setminus E$, where $U$ is the universal set containing all $|V|(|V|-1)/2$ potential links. The task of link prediction is to find out those missing or future links. To test the algorithm's accuracy, the observed links, $E$, is divided into two parts: the training set $E^T$ is treated as known information, while the probe set $E^P$ is used for algorithm evaluation and no information in $E^P$ is allowed to be used for prediction. The majority of known studies applied *random division*, namely $E^P$ is randomly drawn from $E$. In the case of predicting future links, *temporal division* is usually adopted where $E^P$ contains most recently appeared links [16]. In some real networks, missing links have different topological features from observed links. For example, missing links are more likely to be associated with low-degree nodes since interactions between hubs are easy to be known. In such situation, we may apply *biased division* where $E^P$ contains links likely to be similar to missing links [17].

Performance evaluation metrics can be roughly divided into two categories: threshold-dependent metrics (e.g., fixed-threshold accuracy) and threshold-independent metric (e.g., area under threshold curve). Precision and Recall are the two most widely used metrics in the former category. Precision is defined as the ratio of relevant items selected to the number of items selected. That is to say, if we take the top-$L$ links as the predicted ones, among which $L_r$ links are correctly predicted, then the Precision equals $L_r/L$. Recall is defined as the ratio of relevant items selected to the total number of relevant items, say $L_r/|E^P|$. An obvious drawback of threshold-dependent metrics is that we generally do not have a reasonable way to determine the threshold, like the number of predicted links $L$ or the threshold score for the existence of links. A widely adopted way is setting $L = |E^P|$, at which Precision=Recall [3,12]. Although $|E^P|$ is generally unknown, an experiential and reasonable setting is $|E^P| = 0.1|E|$ because 10% of links in the probe set are usually enough for us to get statistical solid results while the removal of 10% of links will probably not destroy the structural features of the target network [18].

Some studies argued that a single value might not well reflect the performance of a predictor [16,19]. Therefore, robust evaluation based on threshold-dependent metrics should cover a range of thresholds (e.g., by varying $L$), which is actually close to the consideration of threshold curves. The Precision-Recall (PR) curve [19] and receiver operating characteristic (ROC) curve [20] are frequently considered in the literature. The former shows Precision with respect to Recall at all thresholds and the latter represents performance trade-off between true positives and false positives at different thresholds. We say algorithm *X* is strictly better than algorithm *Y* only if *X*'s threshold curve completely dominates *Y*'s curve (mathematically speaking, a curve A dominates another curve B in a space is B is always equal or below curve A [21]). Davis and Goadrich [22] have proved that a curve dominates in ROC space if and only if it dominates in PR space.

As the condition of domination is too rigid, we usually use areas under threshold curves as evaluation metrics. The area under the ROC curve (AUC) can be interpreted as the probability that a randomly chosen link in $E^P$ is assigned a higher existence likelihood than a randomly chosen link in $U \setminus E$. If all likelihoods are generated from an independent and identical distribution, the AUC value should be about 0.5. Therefore, the degree to which the value exceeds 0.5 indicates how better the algorithm performs than pure chance. Thus far, AUC is the most frequently used metric in link prediction, probably because it is highly interpretable, easy to be interpolated, and of good visualization. Meanwhile, readers should be aware of some remarkable disadvantages of AUC, for example, AUC is inadequate to evaluate the early retrieval performance which is critical in real applications especially for many biological scenarios [23], and AUC will give misleadingly overhigh score to algorithms that can successfully rank many negatives in the bottom while this ability is less significant in imbalanced learning [19,24]. A typical viewpoint in the early studies is that AUC is suitable for imbalanced learning because AUC is not sensitive to the ratio of positives to negatives and thus can reflect an algorihtm's ability that is independent to the data distribution [25]. However, recent perspective in machine learning and big data is that talking about the performance or ability of a classification algorithm without specified data sets is meaningless, so that this previous advantage gets increasing criticisms [19,24]. Hand argued that AUC is fundamentally incoherent because AUC uses different misclassification cost distributions for different classifiers [26]. Hand's criticism is deep, but AUC indeed measures relative ranks instead of absolute loss and is irrelevant to misclassification cost, so that to dissect AUC in the narration involving misclassification cost may be unfair. To overcome the above disadvantages, scientists have proposed a number of alternatives of AUC, such as H measure [26], concentrated ROC [27] and normalized Discounted Cumulative Gain [28]. At the same time, the area under the Precision-Recall curve (AUPR) becomes increasingly popular, especially for biological studies. Though the AUPR score is less interpretable than AUC, each point in the PR curve has an explicit meaning, and the absolute accuracy metrics (e.g., Precision and Recall) are usually closer to practical requirements than relative ranks.

In summary, empirical comparisons and systematic analyses about evaluation metrics for link prediction are important task in this stage because many publications use AUC as the sole metric, while an ongoing empirical study (by Yan-Li Lee and me, unpublished) shows that in about 1/3 pairwise comparisons AUC and AUPR give different ranks of algorithms, and a recent large-scale experimental study also show inconsistent results by AUC and AUPR [29]. Before a comprehensive and explicit picture obtained, my suggestion is that we have to at least simultaneously report both AUC and AUPR, and only if an algorithm can obviously beat another one in both metrics for a network, we can say the former performs better in this case.

## LOCAL SIMILARITY INDICES

A similarity-based algorithm will assign a similarity score to each nonobserved link, and the one with a higher score is of a larger likelihood to be a missing link. Liben-Nowell and Kleinberg [12] indicated that a very simple index named *common neighbor* (CN), say

$$S_{xy}^{CN} = |\Gamma_x \cap \Gamma_y|, \tag{1}$$

with $\Gamma_x$ and $\Gamma_y$ being sets of neighbors of nodes $x$ and $y$, performs very well in link prediction for social networks. Zhou *et al*. [14] proposed the *resource allocation* (RA) index via weakening the weights of large-degree common neighbors, namely

$$S_{xy}^{RA} = \sum_{z \in \Gamma_x \cap \Gamma_y} \frac{1}{k_z}, \tag{2}$$

where $k_z$ is the degree of node $z$. The simplicity, elegance and good performance of CN, RA and some other alternatives [3] lead to increasing attention on local similarity indices.

In the recent decade, probably the most impressive achievement on local similarity indices is the proposal of the local community paradigm [30], which suggests that two nodes are more likely to link together if their common neighbors are densely connected. Accordingly, Cannistraci *et al*. [30] proposed the CAR index where the CN index is multiplied by the number of observed links between common neighbors, as

$$S_{xy}^{CAR} = S_{xy}^{CN} \cdot \sum_{z \in \Gamma_x \cap \Gamma_y} \frac{|\gamma_z|}{2}, \tag{3}$$

where $\gamma_z$ is the subset of $z$'s neighbors that are also common neighbors of $x$ and $y$. Analogously, RA index can be improved by accounting for the local community paradigm as

$$S_{xy}^{CRA} = \sum_{z \in \Gamma_x \cap \Gamma_y} \frac{|\gamma_z|}{k_z}. \tag{4}$$

By integrating the idea of Hebbian learning rule, the above index is further extended and renamed as Cannistraci-Hebb (CH) index [31]

$$S_{xy}^{CH} = \sum_{z \in \Gamma_x \cap \Gamma_y} \frac{1+k_z^{(i)}}{1+k_z^{(e)}}, \tag{5}$$

where $k_z^{(i)}$ is the internal degree of $z$, say the number of $z$'s neighbors that are also in $\Gamma_x \cap \Gamma_y$, and $k_z^{(e)}$ is the external degree of $z$, say the number of $z$'s neighbors that are not in $\Gamma_x \cap \Gamma_y \cap \{x, y\}$. The core idea of CH index is to consider the negative impacts of external local-community-links (see [32] for more CH indices according to the core idea). Extensive empirical analyses [32][33] indicated that the introduction of local community paradigm and Hebbian learning rule could considerably improve the performance of routine local similarity indices.

In most known studies, the presence of many 2-hop paths between a pair of nodes is considered to be the strongest evidence indicating the existence of a corresponding missing link or future link. Although in local path index [34] and Katz index [35] longer paths are taken into account, they are considered to be less significant than 2-hop paths. Surprisingly, some recent works have argued that 3-hop-based similarity indices perform better than 2-hop-based indices. Pech *et al*. [36] assumed that the existence likelihood of a link is a linear sum of all its neighbors' contributions. After some algebra, Pech *et al*. [36] obtained a global similarity index called linear optimization (LO) index, as

$$S^{LO} = \alpha A(\alpha A^T A + I)^{-1} A^T A = \alpha A^3 - \alpha^2 A^5 + \alpha^3 A^7 - \alpha^4 A^9 + \cdots, \qquad (6)$$

where $A$ and $I$ are adjacency matrix and identity matrix. Clearly, the number of 3-hop paths $A^3$ can be interpreted as a degenerated index of LO. Indeed, Daminelli *et al*. [37] have already applied 3-hop-based indices to predict missing links in bipartite networks, while almost no one at that time has tried 3-hop-based indices on unipartite networks. Kovács *et al*. [38] noted the bipartite nature of the protein-protein interaction networks (not fully bipartite, but of high bipartivity [39]) and independently proposed a degree-normalized index (called *L3* index) based

on 3-hop paths as

$$S_{xy}^{L3} = \sum_{u,v} \frac{a_{xu}a_{uv}a_{vy}}{\sqrt{k_u k_v}}. \quad (7)$$

They showed its advantage compared with 2-hop-based indices in predicting protein–protein interactions. Muscoloni *et al.* [31] further proposed a theory that generalized 2-hop-based indices to *n*-hop-based indices with *n*>2, and demonstrated the superiority of 3-hop-based indices over 2-hop-based indices on protein–protein interaction networks, world trade networks and food webs. For example, in their framework [31], the *n*-hop-based RA index reads

$$S_{xy}^{RA(n)} = \sum_{z_1,z_2,\cdots,z_{n-1}\in \mathbb{L}(n)} \frac{1}{\sqrt[n-1]{k_{z_1}k_{z_2}\cdots k_{z_{n-1}}}}, \quad (8)$$

where $\mathbb{L}(n)$ is the set of all *n*-hop simple paths connecting *x* and *y*, and $z_1, z_2, \cdots, z_{n-1}$ are the intermediate nodes on the considered path. Accordingly, the *L*3 index is exactly the same to the 3-hop-based RA index.

We have implemented extensive experiments on 137 real networks [33], suggesting that: (i) 3-hop-based indices outperform 2-hop-based indices subject to AUC, whilst 3-hop-based and 2-hop-based indices are competitive on Precision; (ii) CH indices performs the best among all considered candidates; (iii) 3-hop-based indices are more suitable for disassortative networks with lower densities and lower average clustering coefficient. Furthermore, we have showed that a hybrid of 2-hop-based and 3-hop-based indices via collaborative filtering techniques can result in overall better performance [40].

# LINK PREDICTABILITY

Quantifying link predictability of a network allows us to evaluate link prediction algorithms for this network, and to see whether there is still a large space to improve the current prediction accuracy. Lü *et al.* [18] raised a hypothesis that missing links are difficult to predict if their addition causes huge structural changes, and thus a network is highly predictable if the removal or addition of a set of randomly selected links does not significantly change structural features of this network. Denote $A$ the adjacency matrix of a simple network $G(V,E)$, and $\Delta A$ the adjacency matrix corresponding to a set of randomly selected links $\Delta E$ from $E$. After the removal of $\Delta E$, the remaining network $G^R$ is also a simple network, so that the corresponding adjacency matrix, $A^R = A - \Delta A$, can be diagonalized as

$$A^R = \sum_{k=1}^{N} \lambda_k x_k x_k^T, \qquad (9)$$

where $N = |V|$, and $\lambda_k$ and $x_k$ are the *k*th eigenvalue and corresponding orthogonal and normalized eigenvector of $A^R$. Considering $\Delta E$ as a perturbation to $A^R$, which results in an updated eigenvalue $\lambda_k + \Delta \lambda_k$ and a corresponding eigenvector $x_k + \Delta x_k$, then we have

$$(A^R + \Delta A)(x_k + \Delta x_k) = (\lambda_k + \Delta \lambda_k)(x_k + \Delta x_k). \qquad (10)$$

Similar to the process to get the expectation value of the first-order perturbation Hamilonian, we neglect the second-order small terms and the changes of eigenvectors, and then obtain

$$\Delta \lambda_k \approx \frac{x_k^T \Delta A x_k}{x_k^T x_k}, \qquad (11)$$

as well as the perturbed matrix

$$\widetilde{A} = \sum_{k=1}^{N} (\lambda_k + \Delta \lambda_k) x_k x_k^T, \qquad (12)$$

which can be considered as the linear approximation of $A$ if the expansion is only based on $A^R$. If the perturbation does not significantly change the structural features, the eigenvectors of $A^R$ and those of $A$ should be almost the same, and thus $\widetilde{A}$ should be very close to $A$ according to Eq. (12). We rank all links in $U \setminus E^R$ in a descending order according to their values in $\widetilde{A}$ and select the

top-$L$ links to form the set $E^L$, where $L = |\Delta E|$. Links in $E^R$ and $E^L$ constitute the perturbed network, and if this network is close to $G$ (because $\widetilde{A}$ is close to $A$), $E^L$ should be close to $\Delta E$. Therefore, Lü *et al.* [18] finally proposed an index called structural consistency to measure the inherent difficulty in link prediction as

$$\sigma_c = \frac{|E^L \cap \Delta E|}{|\Delta E|}. \tag{13}$$

The above perturbation method can also be applied to predict missing links, and the resulted *structural perturbation method* (SPM) is still one of the most accuracy methods till far [29].

Koutra *et al.* [41] found that the major part of a seemingly complicated real network can be represented by a few elemental substructures like cliques, stars, chains, bipartite cores, and so on. Inspired by this study, Xian *et al.* [42] claimed that a network is more regular and thus more predictable if it can be well represented by a small number of subnetworks. To reduce the tremendous complexity caused by countless subnetworks, they further set a strong restriction that candidate subnetworks are ego networks of all nodes, where ego network (also called egocentric network) is a subnetwork induced by a central node (known as the ego) and all other nodes directly connected to the ego (called alters), see [43] for example. Obviously, the ego network of node $i$ can be represented by the $i$th row or $i$th column of the adjacency matrix $A$, and if a network can be perfectly represented by all ego networks, there exists a matrix $Z \in \mathbb{R}^{N \times N}$ such that $A=AZ$. Intuitively, if a network $G$ is very regular, the corresponding representation $Z$ should have three properties: (i) $G$ can be well represented by its ego networks, so that $AZ$ is close to $A$; (ii) $G$ can be well represented by a small number of ego networks, so that $Z$ is of low rank since the redundant ego networks correspond to zero rows in $Z$; (iii) Each ego network of $G$ can be represented by a very few other ego networks, so that $Z$ is sparse. Accordingly, the best representation matrix $Z^*$ can be obtained by solving the following optimization problem

$$\min_{Z} \text{rank}(Z) + \alpha \|Z\| + \beta \|A - AZ\|, \quad (14)$$

where $\alpha$ and $\beta$ are tradeoff parameters. Based on $Z^*$, Xian et al. [42] proposed an ad hoc index named structural regularity index, as

$$\sigma_r = \frac{1}{\sqrt{\frac{n-r}{n}} \sqrt{\frac{\tau}{nr}}}, \quad (15)$$

where $r$ is the rank of $Z^*$, $\tau$ is the number of zero entries in $Z^*$, $\frac{n-r}{n}$ denotes the proportion of identical ego networks, and $\frac{\tau}{nr}$ characterizes the density of zero entries of the reduced echelon form of $Z^*$. Clearly, a lower $r$ and a larger $\tau$ will result in a smaller $\sigma_r$, corresponding to a more predictable network.

Xian et al. [42] suggested that the structural regularity corresponds to redundant information in the adjacency matrix, which can be characterized by a low-rank and sparse representation matrix. Sun et al. [44] proposed a more direct method to measure such redundancy. Their train of thought is that a more predictable network contains more structural redundancy, and thus can be compressed by a shorter binary string. As the shortest possible compression length can be calculated by a lossless compression algorithm, they used the obtained normalized shortest compression length of a network to quantify its structure predictability.

To validate their methods, Lü et al. [18] and Xian et al. [42] tested on many real networks about whether $\sigma_c$ and $\sigma_r$ are strongly correlated with prediction accuracies of a few well-known algorithms. This is rough because any algorithm cannot stand for the theoretical best algorithm.

Sun *et al.* [44] adopted an improved method that uses the best performance among a number of known algorithms for each tested network to approximate the performance of the theoretically best predicting algorithm. Garcia-Perez *et al.* [45] analyzed the ensemble of simple networks, where each can be constructed by generating a link between any node pair *i* and *j* with a known linking probability $p_{ij}$. For such theoretical benchmark, the best-possible algorithm is to rank unobserved links with largest linking probabilities in the top positions and the theoretical limitation of Precision can be easily obtained. They showed that if the size of $\Delta A$ is too small in compared with *A*, the evaluation of predictability by $\sigma_c$ is less accurate.

Structural consistency, structural regularity and compression length are all ad hoc methods. They can be used to probe the intrinsic difficulty in link prediction, but cannot mathematically formulate the limitation of prediction. Mathematically speaking, there could be a God algorithm that correctly predicts all missing links, except for those indistinguishable from nonexistent links. A link (*i*, *j*) is indistinguishable from another link (*u*, *v*) if and only if there is a certain automorphism *f* such that *f(i)=u* and *f(j)=v*, or *f(i)=v* and *f(j)=*u. This extremely rigid definition from automorphism-based symmetry makes virtually all real networks have predictability very close to 1, which is indeed meaningless in practice. Using synthetic networks with known prediction limitation is a potentially promising way to evaluate predictors as well as indices for predictability [45][46], but the results may be irrelevant to real-world networks.

All above studies target static networks, while a considerable fraction of real networks are time-varying (named as temporal networks) [47]. Temporal networks are usually more predictable since one can utilize both topological and temporal patterns. Ignoring topological correlations, the randomness and thus predictability of a time series can be quantified by the entropy rate [48]. Tang *et al.* [49] listed weights of all possible links as an expanded vector with dimension $N^2$ (self-connections are allowed, directionalities are considered), and thus the evolution of a temporal network can be fully described by a matrix $M \in \mathbb{R}^{N^2 \times T}$, where $T$ is the number of snapshots under consideration. After $M$, the evolution of a temporal network can be treated as a stochastic vector process, and how to measure the predictability of temporal networks is transformed to a solved problem based on the generalized Lempel-Ziv algorithm [50]. An obvious defect is that the vector dimension is too big, resulting in huge computational complexity. Tang *et al.* [49] thus replaced $M$ by a much smaller matrix where only links occurring $\geq 10\%$ of snapshots are taken into consideration. An intrinsic weakness of Lempel-Ziv algorithm is that it tends to overestimate the predictability and thus in many situations the estimated values are very close to 1 [48][51]. Tang *et al.* [49] proposed a clever method that compares the predictability of the target network with the corresponding null network, and thus the normalized predictability is able to characterize the topological-temporal regularity in addition to the least predictable one.

## NETWORK EMBEDDING

A network embedding algorithm will produce a function $g: V \to \mathbb{R}^d$ with $d \ll N$, so that every node is represented by a low-dimensional vector [52]. Then, the existence likelihood of a nonobserved link $(u, v)$ can be estimated by the inner product, the cosine similarity, the Euclidean distance or the geometrical shortest path of the two learned vectors $g(u)$ and $g(v)$ [52][53]. Early methods cannot handle large-scale networks because they usually rely on solving the leading eigenvectors [54][55].

Mikolov *et al.* [56][57] proposed a language embedding algorithm named SkipGram that represents every word in a given vocabulary by a low-dimensional vector. Such representation can be obtained by maximizing the co-occurrence probability among words appearing within a window *t* in a sentence, via some stochastic gradient descent methods. Based on SkipGram, Perozzi *et al.* [58] proposed the so-called DeepWalk algorithm, where nodes and truncated random walks are treated as words and sentences. Grover and Leskovec [59] proposed the node2vec algorithm that learns the low-dimensional representation by maximizing the likelihood of preserving neighborhoods of nodes. Grover and Leskovec argued that the choice of neighborhoods play a critical role in determining the quality of the representation. Therefore, instead of simple definitions of the neighborhood of an arbitrary node *u*, such as nodes with distance no more than a threshold to *u* (like the breadth-first search) and nodes sampled from a random walk starting from *u* (like the depth-first search), they utilized a flexible neighborhood sampling strategy by biased random walks, which smoothly interpolates between breadth-first search and depth-first search. Considering a random walk that just traversed link (*z*, *v*) and now resides at node *v*, the transition probability from *v* to any *v*'s immediate neighbor *x* is $\pi_{vx} / \sum_{y \in \Gamma_v} \pi_{vy}$. The node2vec algorithm sets the unnormalized probability as

$$\pi_{vx} = \alpha_{pq}(z, x) \cdot w_{vx}, \tag{16}$$

where $w_{vx}$ is the weight of link (*v*, *x*), and

$$\alpha_{pq}(z, x) = \begin{cases} 1/p, & d_{zx} = 0 \\ 1, & d_{zx} = 1 \\ 1/q, & d_{zx} = 2 \end{cases}. \tag{17}$$

Obviously, the sampling strategy in DeepWalk is a special case of node2vec with *p*=1 and *q*=1. By tuning *p* and *q*, node2vec can achieve better performance than DeepWalk in link prediction.

Tang *et al.* [60] argued that DeepWalk lacks a clear objective function tailored for network embedding, and proposed the LINE algorithm that learns node representations on the basis of a carefully designed objective function that preserves both the first-order and second-order proximity. The first-order proximity is captured by the observed links, and thus can be formulated as

$$O_1 = -\sum_{(u,v)\in E} w_{uv} \log p_1(u,v), \tag{18}$$

where $w_{uv}$ is the weight of the observed link $(u, v)$ and

$$p_1(u,v) = \frac{1}{1+\exp[-g(u)\cdot g(v)]} \tag{19}$$

describes the likelihood of the existence of $(u, v)$ given the embedding $g$. Of course, one can adopt other alternatives of Eq. (19). The second-order proximity assumes that nodes sharing many connections to other nodes are similar to each other. Accordingly, each node is also treated as a specific context and nodes with similar distributions over contexts are assumed to be similar. Then, the second-order proximity can be characterized by the objective function

$$O_2 = -\sum_{(u,v)\in E} w_{uv} \log p_2(u|v), \tag{20}$$

where $p_2(u|v)$ denotes the probability that node *v* will generate a context *u*, namely

$$p_2(u|v) = \frac{\exp[g'(u)\cdot g(v)]}{\sum_{z\in V} \exp[g'(z)\cdot g(v)]}, \tag{21}$$

with $g'(u)$ being the context representation of *u*. Clearly, $O_2$ is naturally suitable for directed networks. By minimizing $O_1$ and $O_2$, LINE learns two kinds of node representations that respectively preserve the first-order and second-order proximity, and takes their concatenation as the final representation.

In addition to DeepWalk, LINE and node2vec, other well-known network embedding algorithms that have been applied in link prediction include DNGR [61], SDNE [62], HOPE [63], GraphGAN [64], and so on. On the one hand, embedding is currently a very hot topic in network science and thought to be a promising method for link prediction. On the other hand, some very recent empirical studies [32][65][66] involving more than a thousand networks showed negative evidence that network embedding algorithms perform worse than some elaborately designed mechanistic algorithms. This is not a bad news because one can expect that some link prediction algorithms will enlighten and energize researchers in network embedding, and thus make contributions to other aspects of network analyses like community detection, classification and visualization.

Another notable embedding method is based on the hyperbolic network model [67][68], where each node is represented by only two coordinates (i.e., $d=2$) in a hyperbolic disk. The hyperbolic network models are very simple yet can reproduce many topological characteristics of real networks, such as sparsity, scale-free degree distribution, clustering, small-world property, community structure, self-similarity, and so on [67][68][69]. Papadopoulos *et al.* [70] applied the hyperbolic model to the Internet at Automomous Systems (AS) level, showed better performance than traditional methods (e.g., CN index and Katz index) in predicting missing links associated with low-degree nodes (see again [17] for biased sampling preferring low-degree nodes). Wang *et al.* [71] proposed an link prediction algorithm for networks with community structure based on hyperbolic embedding, showing good performance for community networks with power-law degree distributions. Alessandro and Cannistraci proposed the so-called nonuniform popularity-similarity-optimization (nPSO) model as a generative model to grow random networks embedded in the hyperbolic space [69], and they leveraged the nPSO model as a synthetic benchmark for link prediction algorithms, showing that the algorithm only accounting for hyperbolic distance does not perform well at the presence of communities [46]. They further proposed a variant of geometric embedding named minimum curvilinear automata (MCA) [72], whose link prediction accuracy is higher than the simple hyperbolic distance [46] but lower than coalescent embedding in hyperbolic space [73]. Kitsak *et al.* [74] proposed an embedding algorithm named HYPERLINK in hyperbolic space, whose goal is to obtain more accurate prediction of missing links, so that HYPERLINK performs better than previous hyperbolic embedding algorithms (most of these algorithms are designed to reproduce topological features, not to predict missing links). HYPERLINK is often competitive to other well-known link predictors and it is in particular good at predicting missing links that are really hard to predict.

## MATRIX COMPLETION

Matrix completion aims to reconstruct a target matrix, given a subset of known entries. Since links can be fully conveyed by the adjacency matrix $A$, it is natural to regard link prediction as a matrix completion task. Denote $E^k$ the set of node pairs corresponding to known entries in $A$ that can be utilized in the matrix completion task. In most studies, $E^k = E^T$, while we should be aware of that $E^k$ can also contain some known absent links. The matrix completion problem can be formulated in line with supervised learning, as

$$\min_{\vartheta} \frac{1}{|E^k|} \sum_{(i,j) \in E^k} \ell[a_{ij}, \tilde{a}_{ij}(\vartheta)] + \Omega(\vartheta), \tag{22}$$

where $\vartheta$ is the parameter vector, $\tilde{a}_{ij}$ is the predicted value of the model, $\ell(\cdot,\cdot)$ is a loss function, and $\Omega$ is a regularization term preventing overfitting. The most frequently used loss functions are squared loss $\ell(a,b)=(a-b)^2$ and logistic loss $\ell(a,b)=\log(1+e^{-ab})$.

Matrix factorization is a very popular method for matrix completion, which has already achieved great success in a closely related domain, the design of recommender systems [75]. We consider a simple network with symmetry $A$, and assume $\tilde{A}$ can be approximately factorized as $\tilde{A} \approx UU^T$ with $U \in \mathbb{R}^{N \times d}$ and $d \ll N$, then we need to solve the following optimization problem

$$\min_U \frac{1}{|E^k|}\sum_{(i,j)\in E^k} \ell\left(a_{ij}, u_i^T u_j\right) + \Omega(U), \tag{23}$$

where $u_i$ and $u_j$ are the $i$th and $j$th rows of $U$. Notice that, $u_i^T$ is the transpose of $u_i$, not the $i$th row of $U^T$. Though without topological interpretation, $u_i$ can be treated as a lower-dimensional representation of node $i$, and matrix factorization can also be considered as a kind of matrix embedding algorithms [76]. If we adopt the squared loss function and the Forbenius norm for $\Omega$, then we get a specific optimization problem

$$\min_U \frac{1}{|E^k|}\sum_{(i,j)\in E^k} \left(a_{ij} - u_i^T u_j\right)^2 + \lambda \|U\|_F^2, \tag{24}$$

where $\lambda$ is a tradeoff parameter. Menon and Elkan [577 suggested that one can directly optimize for AUC on the training set, given some known absent links. Accordingly, the objective function Eq. (23) can be rewritten in terms of AUC as

$$\min_U \frac{1}{|E_+^k||E_-^k|}\sum_{(i,j)\in E_+^k,\,(x,y)\in E_-^k} \ell\left(1, u_i^T u_j - u_x^T u_y\right) + \Omega(U), \tag{25}$$

where $E_+^k$ and $E_-^k$ are sets of known present and known absent links, respectively. Cleary, $E_+^k \cap E_-^k = \emptyset$ and $E_+^k \cup E_-^k = E^k$. Menon and Elkan [77] showed that the usage of AUC-based loss function can improve AUC value by around 10% comparing with the routine loss function like Eq.

(24).

The factorization in Eq. (24) is easy to be extended to directed networks [77], bipartite networks [78], temporal networks [79], and so on. For example, if $A$ is asymmetry, we can replace $\tilde{A} \approx UU^T$ by $\tilde{A} \approx U\Lambda U^T$ with $\Lambda \in \mathbb{R}^{d \times d}$, and thus Eq. (24) can be extended to

$$\min_{U,\Lambda} \frac{1}{|E^k|} \sum_{(i,j) \in E^k} \left(a_{ij} - u_i^T \Lambda u_j\right)^2 + \frac{\lambda_U}{2} \|U\|_F^2 + \frac{\lambda_\Lambda}{2} \|\Lambda\|_F^2, \tag{26}$$

Analogously, for bipartite networks like gene-disease associations [78] and user-product purchases [80], if $A \in \mathbb{R}^{M \times N}$, then we can replace $\tilde{A} \approx UU^T$ by $\tilde{A} \approx WH^T$, where $W \in \mathbb{R}^{M \times d}$ and $H \in \mathbb{R}^{N \times d}$. Accordingly, we get the optimization problem for bipartite networks as

$$\min_{W,H} \frac{1}{|E^k|} \sum_{(i,j) \in E^k} \left(a_{ij} - w_i^T h_j\right)^2 + \frac{\lambda_W}{2} \|W\|_F^2 + \frac{\lambda_H}{2} \|H\|_F^2, \tag{27}$$

where $w_i$ and $h_j$ are the $i$th and $j$th rows of $W$ and $H$, respectively. More details can be found in Refs. [77][78][79].

The explicit features of nodes, such as tags associated with users and products [81], can also be incorporated in the matrix factorization framework. Menon and Elkan [77] suggested a direct combination of explicit features and latent features learned from the observed topology. Denoting $x_i \in \mathbb{R}^s$ the vector of explicit features of node $i$, the predicted values $\tilde{a}_{ij}$ in Eq. (22) are then replaced by

$$\tilde{a}_{ij} = u_i^T u_j + v^T x_i + v^T x_j, \tag{28}$$

where $v \in \mathbb{R}^s$ is a vector of parameters. Experiments showed that the incorporation can considerably improve the prediction accuracy [77]. Jain and Dhillon [82] proposed a so-called inductive matrix completion (IMC) algorithm that uses explicit features to reduce the computational complexity. In IMC, the predicted value can be expressed as $\tilde{a}_{ij} = x_i^T Q Q^T x_j$, where $x_i \in \mathbb{R}^s$ is the vector of $i$'s explicit features, and $Q \in \mathbb{R}^{s \times t}$ is a low-rank matrix with small $t$, which describes the latent relationships between explicit features and topological structure. $Q$ can be learned from observed links by the following optimization problem

$$\min_{Q} \frac{1}{|E^k|} \sum_{(i,j) \in E^k} \left( a_{ij} - x_i^T Q Q^T x_j \right)^2 + \lambda \|Q\|_F^2. \tag{29}$$

Notice that, in Eq. (24), the number of parameters to be learned is $Nd$, while in Eq. (29), we only need to learn $st$ parameters. The numbers of latent features ($d$ and $t$) could be more or less the same as they are largely dependent on the topological structure, while the number of nodes $N$ is usually much larger than the number of explicit features $s$. Therefore, the computational complexity of IMC should be much lower than brute-force factorization methods. The original IMC is proposed for bipartite networks [82], which has already found successful applications in the design of recommender systems [82] and the prediction of gene- and RNA-disease associations [78][83][84].

Pech *et al.* [85] argued that low rank is the most critical property in matrix completion. They assumed that the observed network can be decomposed into two parts as $A = A_B + A_E$, where $A_B$ is called the backbone preserving the network organization pattern, and $A_E$ is a noise matrix, in which positive and negative entries are spurious and missing links, respectively. Pech *et al.* [85] considered only two simple properties: the low rank of $A_B$ and the sparsity of $A_E$. Accordingly, $A_B$ and $A_E$ can be determined by solving the following optimization problem

$$\min_{A_B, A_E} \text{rank}(A_B) + \lambda \|A_E\|_0 \quad \text{subject to} \quad A = A_B + A_E, \tag{30}$$

where $\|A_E\|_0$ is the $l_0$-norm counting the number of nonzero entries of $A_E$. The predicted links can be obtained by sorting entries in $A_B$ that correspond to zero entries in $A$. This method is straightforwardly named as low rank (LR) algorithm. Although being simple, LR performs better than well-known similarity-based algorithms reported in Refs. [14][30], hierarchical structure model [13] and stochastic block model [15], while slightly worse than LOOP [86], structural perturbation model [18], and Cannistraci-Hebb automata [32].

# ENSEMBLE LEARNING

In an early survey [3], we noticed the low stability of individual link predictors and thus suggested ensemble learning as a powerful tool to integrate them. Ensemble learning is a popular method in machine learning, which constructs and integrates a number of individual predictors to achieve better algorithmic performance [87]. Roughly speaking, ensemble learning techniques can be divided into two classes: the *parallel ensemble* where individual predictors do not strongly depend on each other and can be implemented simultaneously (e.g., bagging [88] and random forests [89]) and the *sequential ensemble* where the integration of individual predictors has to be implemented in a sequential way (e.g., boosting [90] and stacking [91]). In the following, we will respectively introduce how parallel ensemble and sequential ensemble are applied to link prediction.

Given an observed network, an individual link predictor will produce a rank of all unobserved links. Pujari and Kanawati [92] proposed an aggregation approach on ranks resulted from individual algorithms. If there are $R$ ranks produced by $R$ individual predictors, an unobserved link $e$'s Borda count $B_k(e)$ in the $k$th rank can be defined as the number of links ranked ahead of $e$ (there are many variants of Borda count and here we use the simplest one). Pujaji and Kanawati

used a weighted aggregation to obtain the final score of any unobserved link $e$, as

$$B(e) = \sum_{k=1}^{R} w_k B_k(e), \tag{31}$$

where $w_k$ is set to be proportional to the precision of the $k$th predictor trained by the observed network. Clearly, smaller $B(e)$ indicates higher existence likelihood. In addition to rank aggregation, similar weighting technique can also be applied in integrating likelihood scores. If every unobserved link is assigned a score (higher score indicates higher existence likelihood) by each predictor, then the final score of any unobserved link $e$ can be defined in a weighted form as

$$S(e) = \sum_{k=1}^{R} w_k S_k(e), \tag{32}$$

where $S_k(e)$ is the score from the $k$th predictor. Different from rank aggregation, $S_k(\cdot)$ ($k = 1, 2, \cdots, R$) should be normalized before the weighted sum to ensure scores from different predictors are comparable. An alternative aggregation method is the ordered weighted averaging (OWA) [93], where the $R$ predictors are ordered according to their importance to the final prediction, as $S^{(1)}, S^{(2)}, \cdots, S^{(R)}$, and then the final score of any unobserved link $e$ is

$$S(e) = \sum_{k=1}^{R} w_k S^{(k)}(e), \tag{33}$$

where $\sum_{k=1}^{R} w_k = 1$ and $w_1 \geq w_2 \geq \cdots \geq w_R \geq 0$. Without prior information, the most usual way to determine the weights is using the maximum entropy method, which maximizes $-\sum_{k=1}^{R} w_k \ln w_k$ subject to $\sum_{k=1}^{R} w_k = 1$ and $\eta = \frac{1}{R-1}\sum_{k=1}^{R-1} (R-k)w_k$, where $\eta$ is a tunable parameter measuring the extent to which the ensemble (33) is like an *or* operation. If $\eta$ is very large, then $w_1 \approx 1$ and $w_k \approx 0$ ($k \geq 2$), that is to say, only the first predictor works. He *et al.* [94] applied OWA to aggregate nine local similarity indices. These indices are ordered according to their normalized values (irrelevant to their qualities), which is essentially unreasonable. Therefore, although He *et al.* [94] reported considerable improvement, later experiments [95][96] indicated that the method in [94] does not work well because the position of a predictor is irrelevant to its quality. In contrast, if the order is relevant to the predictors' qualities (e.g., according to their precisions trained by the target network), OWA will bring in remarkable improvement compared with individual predictors [95]. As some link prediction algorithms scale worse than $\mathbb{O}(N)$, Duan *et al.* [97] argued that to solve smaller problems multiple times is more efficient than to solve a single large problem. They considered a latent factor model (similar to the one described by Eq. (23), with complexity $\mathbb{O}(Nd^2)$), and developed several ways for the bagging decomposition, such as bagging with random nodes together with their immediate neighbors and bagging preferring dense components. They showed that those bagging techniques can largely reduce computational complexity withour sacrificing prediction accuracy. Considering the family of stochastic block models [15], Valles-Catala *et al.* [98] showed that the integration (via MCMC sampling according to Bayesian rules) of individually less plausible models can result in higher predictive performance than the single most plausible model.

Boosting is a typical sequential ensemble algorithm that trains a base learner from initial training set and adjusts weights of instances (the wrongly predicted instances will be enhanced while the easy-to-be-predicted instances will lose weights) in the training set to train the next learner. Such operation will continue until reaching some preseted conditions. The most representative boosting algorithm is AdaBoost [90], which is originally designed for binary classification and thus can be directly applied in link prediction. AdaBoost is an additive model as

$$H(x) = \sum_{t=1}^{T} \alpha_t h_t(x), \tag{34}$$

where $h_t$ is the $t$th base learner, $\alpha_t$ is a scalar coefficient, and $T$ is a preseted terminal time. $H$

aims to minimize the expected value of an exponential loss function

$$\ell(H, W) = \mathbb{E}_{x \sim W}\left[e^{-f(x)H(x)}\right], \tag{35}$$

where $f(x)$ denotes the true class of the instance $x$ and $W$ is the original weight distribution. Without specific requirements, we usually set $W(x) = 1/m$ for every instance $x$ where $m$ is the number of instances. We set $W_1 = W$ and learn $h_1$ from $W_1$, and then $\alpha_1$ is determined by minimizing $\ell(\alpha_1 h_1, W_1)$. The weight of an instance $x$ in the second step is updated as

$$W_2(x) = \begin{cases} \frac{1}{Z_2} W_1(x) e^{-\alpha_1}, & h_1 = f(x) \\ \frac{1}{Z_2} W_1(x) e^{\alpha_1}, & h_1 \neq f(x) \end{cases}, \tag{36}$$

where $Z_2$ is the normalization factor. Obviously, if the instance $x$ can be corrected classified, its weight will decrease, otherwise its weight will increase. Such process iterates until reaching the terminal time $T$. When applying AdaBoost in link prediction, we need to be aware of the following three issues. (i) The base learner should be sensitive to $W_t$, so that we cannot use unsupervised algorithms or supervised algorithms insensitive to $W_t$. (ii) In addition to positive instances (observed links), negative instances should be sampled from unobserved links. Though it introduces some noise, the influence is ignorable if the network is sparse. (iii) The negative instances should be undersampled to keep the data balanced. Comar *et al.* [99] proposed the so-called LinkBoost algorithm, which is an extension of AdaBoost to link prediction with a typical matrix factorization model being the base learner. Instead of undersampling negative instances, they suggest a cost-sensitive loss function which penalizes the misclassifying links as nonlinks about $N$ times heavier than misclassifying nonlinks as links. They further considered a degree-sensitive loss function that penalizes more for misclassification of links between low-degree nodes than high-degree nodes.

Stacking [91] is another powerful approach in sequential ensemble. It trains a group of primary learners from the initial training set and uses the outputs of primary learners as input features to train the secondary learner that provides the final prediction. If both primary learners (i.e., input features) and training instances are directly generated by the same training set, the risk of overfitting will be very high. Therefore, the original training set $D$, usually containing similar numbers of positive and negative instances for data balance, is divided into $J$ sets with same size as $D_1, D_2, \cdots, D_J$. Denoting $h_r^{(j)}$ the primary learner using the $r$th algorithm and trained from the $j$th fold of the training set $\overline{D_j} = D \setminus D_j$, for each instance $x_i \in D_j$, its feature vector is $z_i = (z_{i1}, z_{i2}, \cdots, z_{iR})$, where $z_{ir} = h_r^{(j)}(x_i)$ and $R$ is the number of primary algorithms. This $J$-fold division ensures all features of any instance $x$ are obtained by primary learners trained without $x$. Some scientists have already used similar techniques (e.g., using various regressions to integrate results from primary predictors and other features [96][100][101]), but they are not aware of stacking model and did not employ any measures to avoid overfitting. Li *et al.* [102] proposed a stacking model for link prediction, which use logistic regression and XGBoost to learn 4 similarity indices. Their method is inspiring, but they only considered 4 primary predictors and tested on two very small networks with some experimental results (e.g., the AUC values of CN index) far different from well-known results, and thus the reported results and conclusion are questionable. Ghasemian *et al.* [66] proposed a stacking model that considers 203 primary link predictors on 550 disparate networks. Using a standard supervised random forest algorithm [89] as the secondary learner, Ghasemian *et al.* [66] argued that the stacking model is remarkably superior to individual predictors for real networks, and can approach to the theoretical optima for synthetic networks with known highest prediction accuracies. In addition, they showed that social networks are more predictable than biological and technological networks. However, a recent large-scale experiment [29] suggested that the above stacking model does not perform better than SPM [18] and Cannistraci-Hebb automata [32]. Wu *et al.* [77] proposed an alternative sequential ensemble strategy called network reconstruction, which reconstructs network via one link prediction algorithm and predict missing links by another prediction algorithm.

## DISCUSSION

In this review, to improve the readability, we classify representative works in the last decade into five groups. Of course, some novel and interesting methods, such as evolutionary algorithm [103], ant colony approach [104], structural Hamiltonian analysis [86] and leading eigenvector control [105], do not belong to any of the above groups, and readers are encouraged to read other recent surveys [106][107][108] as complements of the present review.

Computational complexities of some representative algorithms mentioned in this Perspective are reported in Table 1. Those algorithms are roughly categorized into three classes according to their computational complexities: the ones of *high* scalability can be applied to large-scale networks with millions of nodes, the ones of *medium* scalability can be applied to mid-sized networks with tens to hundreds of thousands of nodes, and the ones of *low* scalability can only deal with small networks with up to a few thousands of nodes using a common desktop computer. Given the size and sparsity of the target network, as well as the computational power, this table helps readers in finding suitable algorithms.

Table 1: Computational complexities of some representative link prediction algorithms. N is the number of nodes, ⟨k⟩ is the average degree, γ is usually a large number depending on the number of random walks, the length of each walk, and the implemented deep learning model, t is the number of iterations, and d is the representation dimension or the preseted rank for matrix factorization and low-rank decomposition.

| Algorithm | Reference | Complexity | Scalability |
|---|---|---|---|
| CN | [12] | $\mathbb{O}(N\langle k\rangle^2)$ | High |
| RA | [14] | $\mathbb{O}(N\langle k\rangle^2)$ | High |
| LP | [14] | $\mathbb{O}(N\langle k\rangle^3)$ | Medium |
| CH | [31] | $\mathbb{O}(N\langle k\rangle^3)$ | Medium |
| L3 | [38] | $\mathbb{O}(N\langle k\rangle^3)$ | Medium |
| SPM | [18] | $\mathbb{O}(dN^2)$ | Medium |
| DeepWalk | [58] | $\mathbb{O}(\gamma N \log N)$ | Medium |
| LINE | [60] | $\mathbb{O}(tN\langle k\rangle)$ | Medium |
| NetSMF | [76] | $\mathbb{O}(Nd^2)$ | Medium |
| IMC | [82] | $\mathbb{O}(N\langle k\rangle d^2)$ | Medium |
| LR | [85] | $\mathbb{O}(dN^2)$ | Medium |
| HSM | [13] | $\mathbb{O}(e^N)$ | Low |
| SBM | [15] | $\mathbb{O}(e^N)$ | Low |
| LOOP | [86] | $\mathbb{O}(N^3)$ | Low |

Very recently, a notable issue is the applications of neural networks in link prediction, which may be partially facilitated by the dramatic advances of deep learning techniques. Zhang and Chen [109] trained a fully-connected neural network on the adjacency matrices of enclosing subgraphs (with a fixed size) of target links. They applied a variant of the Weisfeiler-Lehman algorithm to determine the order of nodes in each adjacency matrix, ensuring that nodes with closer distances to the target link are ranked in higher positions. Zhang and Chen [110] further proposed a novel framework based on graph neural networks, which can learn multiple types of information, including general structural features and latent and explicit node features. In this framework, a node's order in the enclosing subgraph can be determined only by its closeness to the target link and the subgraph size can be flexible. Wang *et al.* [111] directly represented the adjacency matrix of a network as an image and then learned hierarchical feature representations by training generative adversarial networks. Some preliminary experimental results suggested that the performance of those methods [109][110][111] is highly competitive to many other state-of-the-art algorithms. In despite of the promising results, at present, features and models are simply pieced together without intrinsic connections. The above pioneering works [109][110][111] provide a good start but we still need in-depth and comprehensive analyses to push forward related studies.

Although most link prediction algorithms only account for structural information, attributes of nodes (e.g., expression levels of genes [78] and tags of citation and social networks [81][112]) can be utilized to improve the prediction performance. It is easy to treat attributes as independent information additional to structural features and work out a method that directly combines the two, while what is lacking but valuable is to uncover nontrivial relationship between attributes and structural roles and then design more meaningful algorithms. Beyond explicit attributes, we should also pay attention to dynamical information. It is known to us that limited time series obtained from some dynamical processes can be used to reconstruct network topology [113] while even a small fraction of missing links in modeling dynamical processes can lead to remarkable biases [114], however, studies about how to make use of the correlations between topology and dynamics to predict missing links or how to take advantage of link prediction algorithms to improve estimates of dynamical parameters are rare.

By an elaborately designed model, Gu *et al.* [115] showed that there is no ground truth in ranking influential spreaders even with a given dynamics. Peel *et al.* [116] proved that there is no ground truth and no free lunch for community detection. The latter implies that no detection algorithm can be optimal on all inputs. Fortunately, we have ground truth in link prediction, however, extensive experiments [66] also implicate that no known link predictor performs best or worst across all inputs. If link prediction is a no-free-lunch problem, then no single algorithm performs better or worse than any other when applied to all possible inputs. It raises a question that whether the study on prediction algorithms is valuable. The answer is of course YES [117], because we actually have free lunches as what we are interested in, the real networks, have far different statistics from those of all possible networks. As Ghasemian *et al.* [66] argued that the ensemble models are usually superior to individual algorithms, a related question is whether the study on individual algorithm is valuable. The answer is still YES. Firstly, a recent large-scale experimental study [29] indicated that the performance of the stacking model is worse than elaborately-designed individual algorithms, like SPM [18] and Cannistraci-Hebb automata [32]. Secondly, an individual algorithm could be highly cost-effective for its competitive performance and low complexity in time and space. Above all, individual algorithms, especially the mechanistic algorithms, may provide significant insights about network organization and evolution. In some real applications like friend recommendation, predictions with explanations are more acceptable [118], which cannot be obtained by ensemble learning. In addition, an alogical reason is that some elegant individual models (e.g., HSM [13], SBM [15], SPM [18], HYPERLINK [74], etc.) bring us inimitable aesthetic perception that cannot be experienced elsewhere.

Along with fruitful algorithms proposed recently, the design of novel and effective algorithms for general networks is increasingly hard. We expect a larger fraction of algorithms in the future studies will be designed for networks of particular types (e.g., directed networks [119], weighted networks [120], multilayer networks [121], temporal networks [122], hypergraphs and bipartite networks [37][123], networks with negative links [124][125], etc.) and networks with domain knowledge (e.g., drug-target interactions [126], disease-associated relations [127], protein-protein interactions [38][128], criminal networks [129], citation networks [130], academic social networks [131], knowledge graphs [132], etc.). We should take serious consideration about properties and requirements of target networks and domains in the algorithm design, instead of straightforward (and thus less valuable) extensions of general algorithms. For example, if we attempt to recommend friends in an online social network based on link prediction [10], we need to consider how to explain our recommendations to improve the acceptance rate [118], how to use the acceptance/rejection information to promote the prediction accuracy [133], and how to avoid recommending bots to real users [134]. These considerations will bring fresh challenges in link prediction.

Early studies often compare a very few algorithms on several small networks according to one or two metrics. Recent large-scale experiments [29][32][33][65][66] indicated that the above methodology may result in misleading conclusions. Future studies ought to implement systematic analyses involving more synthetic and real networks, benchmarks, state-of-the-art algorithms and metrics. Researchers can find benchmark data sets for networks from Open Graph Benchmark (OGB, ogb.stanford.edu), Pajek (mrvar.fdv.uni-lj.si/pajek), Link Prediction Benchmarks (LPB,

www.domedata.cn/LPB). If relevant results cannot be published in an article with limited space, they should be made public (better together with data and codes) in some accessible websites like GitHub, OGB and LPB.

Lastly, we would like to emphasize that the soul of a network lies in its links instead of nodes, otherwise we should pay more attention on set theory rather than graph theory. Therefore, in network science, link prediction is a paradigmatic and fundamental problem with long attractivity and vitality. Beyond an algorithm predicting missing and future links, link prediction is also a powerful analyzing tool, which has already been utilized in evaluating and inferring network evolving mechanisms [6][7][100], testing the privacy-protection algorithms (as an attaching method) [135], evaluating and designing network embedding algorithms [136][137], and so on. Though the last decade has witnessed plentiful and substantial achievements, the study of link prediction is just unfolding and more efforts are required towards a full picture of how links do emerge and vanish.

**Acknowledgements**. I acknowledge Yan-Li Lee for valuable discussion and assistance. This work was partially supported by the National Natural Science Foundation of China (Grant Nos. 11975071 and 61673086 ), the Science Strength Promotion Programmer of UESTC under Grant No. Y03111023901014006, and the Fundamental Research Funds for the Central Universities under Grant No. ZYGX2016J196 .